\documentclass[12pt]{article}
\usepackage{amssymb}
\usepackage{epsfig}
\usepackage{psfrag}
\usepackage{latexsym}
\usepackage{euscript}
\usepackage{feynmp}
\usepackage{textcomp}
\def\gs{\mathrel{
   \rlap{\raise 0.511ex \hbox{$>$}}{\lower 0.511ex \hbox{$\sim$}}}}
\def\ls{\mathrel{
   \rlap{\raise 0.511ex \hbox{$<$}}{\lower 0.511ex \hbox{$\sim$}}}}

\newcommand{\ba}{\begin{array}{c}}
\newcommand{\baz}{\begin{array}{cc}}
\newcommand{\bad}{\begin{array}{ccc}}
\newcommand{\ea}{\end{array}}

\newcommand{\be}{\beta}

\def\beq{\begin{equation}}
\def\eeq{\end{equation}}
\def\bea{\begin{eqnarray}}
\def\eea{\end{eqnarray}}
\def\bet{\begin{tabular}}
\def\eet{\end{tabular}}
\def\bes{\begin{subequations}\bea}
\def\ees{\eea\end{subequations}}

\def\be{\begin{equation}}
\def\ee{\end{equation}} 

\def\bea{\begin{eqnarray}}
\def\eea{\end{eqnarray}}
\def\nn{\nonumber}

\def\be{\begin{equation}}
\def\ee{\end{equation}}
\def\bc{\begin{center}}
\def\ec{\end{center}}
\def\bea{\begin{eqnarray}}
\def\eea{\end{eqnarray}}
\def\dd{\displaystyle}
\def\nn{\nonumber}

\catcode`@=11
\def\marginnote#1{}
\newcount\hour
\newcount\minute
\newtoks\amorpm
\hour=\time\divide\hour by60
\minute=\time{\multiply\hour by60 \global\advance\minute by-\hour}
\edef\standardtime{{\ifnum\hour<12 \global\amorpm={am}%
        \else\global\amorpm={pm}\advance\hour by-12 \fi
        \ifnum\hour=0 \hour=12 \fi
        \number\hour:\ifnum\minute<10 0\fi\number\minute\the\amorpm}}
\edef\militarytime{\number\hour:\ifnum\minute<10 0\fi\number\minute}
\def\draftlabel#1{{\@bsphack\if@filesw {\let\thepage\relax
   \xdef\@gtempa{\write\@auxout{\string
      \newlabel{#1}{{\@currentlabel}{\thepage}}}}}\@gtempa
   \if@nobreak \ifvmode\nobreak\fi\fi\fi\@esphack}
        \gdef\@eqnlabel{#1}}
\def\@eqnlabel{}
\def\@vacuum{}
\def\draftmarginnote#1{\marginpar{\raggedright\scriptsize\tt#1}}
\def\draft{\oddsidemargin 0.0truein
        \def\@oddfoot{\sl preliminary draft \hfil
        \rm\thepage\hfil\sl\today\quad\militarytime}
        \let\@evenfoot\@oddfoot \overfullrule 3pt
        \let\label=\draftlabel
        \let\marginnote=\draftmarginnote
   \def\@eqnnum{(\theequation)\rlap{\kern\marginparsep\tt\@eqnlabel}%
\global\let\@eqnlabel\@vacuum}  }
\catcode`@=12

\begin{document}
\begin{titlepage}
\vspace*{-1cm}
\phantom{hep-ph/***}
\hfill{DFPD-10/TH/23}\\
\vskip 2.5cm
\begin{center}
\mathversion{bold}
{\Large\bf The Golden Ratio Prediction for the Solar Angle\\
\vspace{0.5 cm}
from a Natural Model with $A_5$ Flavour Symmetry}
\mathversion{normal}
\end{center}
\vskip 0.5  cm
\begin{center}
{\large Ferruccio Feruglio}~\footnote{e-mail address: feruglio@pd.infn.it} and
{\large Alessio Paris}~\footnote{e-mail address: paris@pd.infn.it}
\\
\vskip .2cm
Dipartimento di Fisica `G.~Galilei', Universit\`a di Padova
\\
INFN, Sezione di Padova, Via Marzolo~8, I-35131 Padua, Italy
\end{center}
\vskip 0.7cm
\begin{abstract}
\noindent
We formulate a consistent model predicting, in the leading order approximation, maximal atmospheric mixing angle,
vanishing reactor angle and $\tan \theta_{12}=1/\phi$ where $\phi=(1+\sqrt{5})/2$ is the Golden Ratio.
The model is based on the flavour symmetry $A_5\times Z_5\times Z_3$, spontaneously broken by a set of flavon fields.
By minimizing the scalar potential of the theory up to the next-to-leading order in the symmetry breaking parameter,
we demonstrate that this mixing pattern is naturally achieved in a finite portion of the parameter space, through the vacuum alignment
of the flavon fields. The leading order approximation is stable against higher-order corrections. We also compare our construction
to other models based on discrete symmetry groups.
\end{abstract}
\end{titlepage}
\setcounter{footnote}{0}
\vskip2truecm
%
%
\section{Introduction}
Neutrino oscillations have revealed unexpected features of the flavour problem. The mixing matrix in the lepton sector is completely different from that in the quark sector and the relative hierarchy between neutrino masses, given in terms of the ratio between the two independent
squared mass differences, is small compared to the hierarchy for charged fermions. The regular pattern observed in the
quark sector strongly suggests that quark masses and mixing angles should be explained by some dynamical principle which, in a unified picture, should eventually apply
also to leptons. In searching for such a common principle a valuable tool is that
of spontaneously broken flavour symmetries, which might be tentatively identified from the present data.

The lepton mixing matrix $U_{PMNS}$ 
still suffers from large uncertainties. The parameters related to CP violations are totally
unknown at present. The reactor angle $\theta_{13}$ is the smallest mixing angle,
but there is only an upper bound on it and its value can range from zero to about 0.2.
The atmospheric mixing angle $\theta_{23}$ is compatible with being maximal, but
deviations from maximality are still allowed to some extent. The most precisely measured angle
is the solar angle $\theta_{12}$, which is large but not maximal, with a 1$\sigma$ uncertainty
of less than 2 degrees:
\be
\sin^2\theta_{12}=0.304^{+0.022}_{-0.016}~~~\cite{schw}~~~,~~~~~~~~~~\sin^2\theta_{12}=0.321^{+0.023}_{-0.022}~~~\cite{gonz}~~~.
\label{exp_12}
\ee
Despite the lack of a precise knowledge of $U_{PMNS}$,
the present data can guide us in searching for a first-order approximation 
providing the basis of a theoretical description. Spontaneously broken flavour symmetries
are suitable to this type of description since they typically generate a mixing matrix 
at the leading order (LO) of the expansion in the symmetry breaking parameter, with small sub-leading
corrections originating from the higher-orders. In the quark sector a natural first approximation is given by $V_{CKM}=1$. 
In the lepton sector the identification of
such a first-order approximation is less obvious and several candidates have been proposed so-far.
Most of them have in common atmospheric and reactor angles, fixed to the values $\sin^2\theta_{23}=1/2$ and $\theta_{13}=0$,
and they only differ in the solar mixing angle. Tri-bimaximal mixing (TB) is perhaps the most studied pattern \cite{TB}. It predicts 
$\sin^2\theta_{12}=1/3$, which is within two standard deviations from the current best value. TB
mixing can be reproduced at the LO in many models based on discrete and continuous flavour symmetries  \cite{review}.
A minimal construction is based on $A_4$. In bimaximal mixing (BM) the solar angle is maximal, $\sin^2\theta_{12}=1/2$, outside the presently 
allowed range \cite{BM, BMmixing}. To reconcile the LO approximation with the data, the expansion parameter should be not-too-small, 
of the order of the Cabibbo angle. Sub-leading corrections of this size are expected to affect also other parameters, 
such as $\theta_{13}$, which is thus predicted close to the present experimental upper bound.

Another plausible mixing pattern is the one where $\sin^2\theta_{23}=1/2$, $\theta_{13}=0$
and $\tan \theta_{12}=1/\phi$ where $\phi=(1+\sqrt{5})/2$ is the Golden Ratio (GR) \cite{other_golden,Strumia:2007}. This pattern,
called GR the hereafter, is the focus of the present paper. We have
\be
\sin^2\theta_{12}=\frac{1}{\sqrt{5}\phi}=\frac{2}{5+\sqrt{5}}\approx 0.276~~~.
\label{gora}
\ee
This value is about two standard deviations below the experimental range and can be brought inside
the allowed interval by a small NLO correction, of order 0.05 radiants. \footnote{An alternative proposal \cite{WR} relating the Golden Ratio to the lepton mixing assumes $\cos\theta_{12}=\phi/2$. Consequently we have
$\sin^2\theta_{12}=\frac{1}{4}(3-\phi)\approx 0.345$, about two standard deviations above the experimental value. In \cite{Rode:2009}
this prediction was deduced from the symmetry of the dihedral group $D_{10}$. }
The GR characterizes several properties of the icosahedron and a natural candidate for the flavour symmetry giving rise to (\ref{gora})
is the icosahedral one, related to the group $A_5$. This relation was pointed out for the first time while trying to 
connect the value of the solar angle to the Golden Ratio \cite{Strumia:2007}.
Indeed there have been attempts to construct a model based on the $A_5$ symmetry \cite{everett:2009} to the purpose of reproducing
the GR mixing pattern, but a complete model does not exist to date. Recently, the group $A_5$ was also applied to a scenario with a fourth
lepton family \cite{Chen:2010}, while the double cover of the icosahedral group was used to reproduce the quark mixing \cite{everett:2010}.
In models based on spontaneously broken flavour symmetries a crucial
feature is the discussion of the vacuum alignment. The family symmetry is broken by the VEVs 
of flavon fields and the desired mixing pattern is intimately related to the directions of these VEVs in flavour space.
In a complete model the VEV alignment should occur naturally, as the outcome of the minimization of 
the energy density of the theory. To our knowledge none of the existing proposals of what we called the GR pattern have solved the vacuum
alignment problem. 

Aim of our work is to build a complete model based on the family group $A_5$
and reproducing the GR mixing pattern in a natural way. In section 2 we will discuss how the symmetry group
$A_5$ can be used to generate the GR mixing pattern. We show that the most general neutrino mass matrix
giving rise to the GR mixing pattern is invariant not only under the $\mu-\tau$ exchange symmetry, but also
under a parity transformation $S$. At the same time the requirement of a diagonal charged lepton mass matrix
is guaranteed by the invariance under a transformation $T$ of order five. We will see that the elements $S$ and $T$ actually 
generate the group $A_5$. Some properties of this group will be recalled in section 3. In section 4 we define our model
by assigning all fields to representations of the flavour group. We also list the flavon VEVs needed in order to accomplish
the desired symmetry breaking and generate the GR ratio. The minimization of the scalar potential is performed
in section 5 and 6, respectively at the LO and NLO. In section 6 we also enumerate the other sources of 
corrections to the LO approximation. We compare our model to other proposals in section 7 and then we conclude.
%
%
\section{A family symmetry for the Golden Ratio}
We start by analyzing the property of the most general neutrino mass matrix leading to the Golden Ratio (GR) prediction for the solar mixing angle.
We chose a basis where the mass matrix for the charged leptons $m_l$ is diagonal. More precisely, it is sufficient that the combination
$m_l^\dagger m_l$ is diagonal, so that there is no contribution to the lepton mixing from the charged lepton sector.
We should also make a choice for $\theta_{23}$ and $\theta_{13}$. To begin with we assume a leading order approximation
where $\sin^2 \theta_{23}=1/2$, $\sin^2 \theta_{13}=0$ and $\tan \theta_{12}=1/\phi$ where $\phi=(1+\sqrt{5})/2$ is the GR.
We look for the most general neutrino mass matrix $m_\nu$ leading to this mixing pattern.
Such a matrix can be constructed by acting with the corresponding mixing matrix $U_{GR}$ on a generic diagonal neutrino mass matrix:
\be
m_{\nu}=U_{GR}^* ~{\tt diag}(m_1,m_2,m_3)~ U_{GR}^\dagger~~~.
\label{numass1}
\ee
In a particular phase convention, the matrix $U_{GR}$ representing our mixing pattern is given by:
\vskip 0.2cm
\begin{equation}
U_{GR}= \left(\matrix{ \cos{\theta_{12}}&\sin{\theta_{12}}&0\cr
& &\\[-0.3cm]\cr
\dd\frac{\sin{\theta_{12}}}{\sqrt 2}&-\dd\frac{ \cos{\theta_{12}}}{\sqrt 2}&\dd\frac{1}{\sqrt 2}\cr
& &\\[-0.1cm]\cr
\dd\frac{\sin{\theta_{12}}}{\sqrt 2}&-\dd\frac{ \cos{\theta_{12}}}{\sqrt 2}&-\dd\frac{1}{\sqrt 2}}\right)~~~,
\label{2.1}
\end{equation}
\vskip 0.5cm\noindent
with $\tan\theta_{12}=1/\phi$. 
By applying eq. (\ref{numass1}) we find a matrix of the form:
\begin{equation}
m_\nu=\left(\matrix{
x&y&y\cr
y&z&w\cr
y&w&z}\right)~~~,
\label{gl1}
\end{equation}
with coefficients $x$, $y$, $z$ and $w$ satisfying the following relation:
\be
z+w=x-\sqrt{2} y~~~.
\label{solarGR}
\ee
The matrix in eq. (\ref{gl1}) is the most general one giving rise to
$\theta_{13}=0$ and $\theta_{23}$ maximal. The constraint of eq. (\ref{solarGR}) arises from further specifying the solar mixing angle.

The matrix in eqs. (\ref{gl1}-\ref{solarGR})  can be completely characterized by a simple symmetry requirement. 
Indeed, it is invariant under the action of the two unitary transformations:
\be
U=\left(
\begin{array}{ccc}
1&0&0\\
0&0&1\\
0&1&0
\end{array}
\right)~~~~~~~~~~~~~
S=\dd\frac{1}{\sqrt{5}} \left(
\begin{array}{ccc}
1&\sqrt{2}&\sqrt{2}\\
\sqrt{2}&-\phi&1/\phi\\
\sqrt{2}&1/\phi&-\phi
\end{array}
\right)~~~,
\label{sandu}
\ee
which satisfy 
\be
S^2=U^2=1~~~,~~~~~~~~~~[S,U]=0~~~,
\ee
and generate a group $G_\nu=Z_2\times Z_2$.
Conversely, the requirement of invariance under $U$ and $S$ completely characterize $m_\nu$ in eqs. (\ref{gl1}-\ref{solarGR}). 
Namely, given a generic neutrino mass matrix $m_{\nu}$ the most general solution to the equations:
\be
U^T~m_{\nu}~U=m_\nu~~~,~~~~~~~S^T~m_{\nu}~S=m_\nu~~~,
\ee
with $U$ and $S$ given in eq. (\ref{sandu}), is the mass matrix defined by eqs. (\ref{gl1}) and (\ref{solarGR}). 

In the chosen basis, where $m_l^\dagger m_l$ is diagonal, there is no contribution to the lepton mixing from the charged lepton sector and the mixing matrix $U_{GR}$ originates
only from the diagonalization of $m_{\nu}$. To construct a model for the desired mixing pattern, we should require that
a diagonal $m_l^\dagger m_l$ arises naturally, as the general solution of a symmetry or dynamical requirement.
For instance, we can require that the charged lepton sector is invariant under a family group $G_l$ enforcing a diagonal $m_l^\dagger m_l$. 
In our LO approximation the groups $G_\nu$ and $G_l$ should be seen as
the residual vacuum symmetries characterizing the neutrino sector and the charged lepton sector,
respectively. Such a configuration can be induced by the spontaneous breaking of some
family symmetry $G_f$, through the vacuum expectation values of two different sets of flavons that selectively couple
to neutrinos and to charged leptons. It is not strictly necessary that $G_f$ entirely contains $G_l$ and $G_\nu$ as subgroups,
since a part of the residual symmetries can arise accidentally, due to the specific field content of the model,
as the baryon and the lepton numbers arise as accidental classical symmetries in the standard model. 
A natural candidate for the family symmetry $G_f$ giving rise to the GR prediction for the solar mixing angle is the proper symmetry group of the icosahedral,
the alternating group $A_5$ \cite{A5}. One of the possible presentations of $A_5$ is in term of two generators $S$ and $T$ satisfying:
\be
S^2=(ST)^3=1~~~~~~~~~~{\rm and}~~~~~~~~~~T^5=1~~~.
\label{present0}
\ee
We make the following ansatz: we identify the matrix $S$ in eq. (\ref{sandu}) with
the generator $S$ of $A_5$. Given the explicit form of the generator $S$, the algebraic relation (\ref{present0}) allows to
determine the matrix corresponding to the generator $T$. We find:
\be
T=\left(
\begin{array}{ccc}
1&0&0\\
0&e^{\dd \frac{2\pi i}{5}}&0\\
0&0&e^{\dd \frac{8\pi i}{5}}
\end{array}
\right)~~~.
\label{t}
\ee
This is an encouraging result. Indeed the condition 
\be
T^\dagger~(m^\dagger_l m_l)~T=(m^\dagger_l m_l)
\ee
requires $m^\dagger_l m_l$ to be a diagonal matrix and the natural candidate for the subgroup $G_l$ is the group $Z_5$ generated by $T$. 
We look for a model invariant under the family symmetry $A_5$, where, after spontaneous breaking,  the residual symmetries of the neutrino 
sector and of the charged lepton sector are those generated by $(S,U)$ and $T$, respectively. It turns out that, up to an irrelevant overall sign,
the $\mu-\tau$ symmetry is an element of $A_5$, that can be expressed in terms of the generators $S$ and $T$ of eqs. (\ref{sandu}) and (\ref{t})
\footnote{We thank Claudia Hagedorn for pointing this out to us.}.
By construction the model predicts GR for the solar mixing angle. We expect that higher order corrections produce small deviations from the LO predictions 
$\theta_{13}=0$ and $\theta_{23}=\pi/4$ and $\tan \theta_{12}=1/\phi$.

Notice that this approach automatically guarantees the independence of the mixing matrix $U_{GR}$ and the other physical results from the base choice.
Indeed, in a generic basis where the generators are $X_\Omega=\Omega~ X~ \Omega^\dagger$ $(X=S,T,U)$, $\Omega$ denoting a unitary
3 $\times$ 3 matrix, in general the combination $m_l^\dagger m_l$ is no more diagonal and the neutrino mass matrix $m_\nu$
have a texture different from the one in eqs. (\ref{gl1}-\ref{solarGR}). However, as a result of the residual symmetries,
$m_l^\dagger m_l$ is diagonalized by $\Omega$, whereas $m_{\nu}$ is diagonalized by $(\Omega~U_{GR})$,
the physical mixing matrix remaining $U_{GR}$.
%
%
\section{The $A_5$ group}
The group $A_5$ is the group of the even permutations of five objects. It is the proper symmetry group of two of the five Platonic solids,
the icosahedron and the dodecahedron. It has 60 elements that can be grouped into five conjugacy classes with 1, 12, 12, 15 and 20 elements.
The five irreducible representations are the invariant singlet, two inequivalent triplets, a tetraplet and a pentaplet. The characters of $A_5$ are collected in Table 1.
\begin{table}[h!]
\begin{center}
\begin{tabular}{c|ccccc}
&   \\
{$ { A_5}$}~~& $C_1$&\hfill $12C^{[5]}_2$& \hfill$12C^{[5]}_3$&\hfill $15C^{[2]}_4$&\hfill $20C^{[3]}_5$\\
 &&&&&   \\
\hline    
 &&&&&   \\
$\chi^{[\bf 1]}_{}$~~& $1$&\hfill $1$&\hfill  $1$&\hfill  $1$&\hfill $1$\\
 &&&&&   \\
$\chi^{[\bf 3]}_{}$~~& $3$&\hfill  $\phi$&\hfill  $(1-\phi)$&\hfill  $-1$&\hfill $0$ \\
 &&&&&   \\
$\chi^{[\bf 3']}_{}$~~& $3$&\hfill  $(1-\phi)$&\hfill  $\phi$&\hfill  $-1$&\hfill $0$\\
 &&&&&   \\
$\chi^{[\bf 4]}_{}$~~& $4$&\hfill  $-1$&\hfill  $-1$&\hfill  $0$&\hfill $1$\\
 &&&&&   \\
$\chi^{[\bf 5]}_{}$~~& $5$& \hfill $0$&\hfill  $0$&\hfill  $1$&\hfill $-1$\\
\end{tabular}
\caption{Characters of the $A_5$ group.}
\end{center}
\label{ca5}
\end{table}
The products of two $A_5$ representations can be decomposed according to the following rules
\bea
3\otimes3&=&(1+5)_S+3_A\nn\\
3'\otimes3'&=&(1+5)_S+3'_A\nn\\
3\otimes3'&=&4+5\nn\\
3\otimes4&=&3'+4+5\nn\\
3'\otimes4&=&3+4+5\nn\\
3\otimes5&=&3+3'+4+5\\
3'\otimes5&=&3+3'+4+5\nn\\
4\otimes4&=&(1+4+5)_S+(3+3')_A\nn\\
4\otimes5&=&3+3'+4+5+5\nn\\
5\otimes5&=&(1+4+5+5)_S+(3+3'+4)_A\nn
\eea
where the suffices $S$($A$) denote the symmetric(antisymmetric) property of the corresponding representation.
The product between the singlet and any representation $r$ gives $r$. As recalled in the previous section, $A_5$
is generated by two elements $S$ and $T$, with the presentation
\be
S^2=(ST)^3=1~~~~~~~~~~{\rm and}~~~~~~~~~~T^5=1~~~.
\label{present1}
\ee
The element $S$ belongs to the class $C_4^{[2]}$ and the element $T$ to the class $C_2^{[5]}$.
We find useful to work in a basis where the $T$ generator for the various representations is always diagonal. Since this choice is unconventional
we list in Table 2 the matrices associated to $S$ and $T$ in our basis. In Appendix B we made connection with other basis used in the literature.
\vskip 0.5cm
\begin{table}[h!]
\begin{center}
\begin{tabular}{|c|c|c|}
\hline
& &\\[-0.3cm]
&$S$&$\dd\frac{5}{2\pi i}\log(T)$\\
& &\\[-0.3cm]
\hline
& &\\[-0.3cm]
$3$&$ \frac{1}{\sqrt{5}}\left(
\begin{array}{ccc}
 1 & \sqrt{2} & \sqrt{2} \\
 \sqrt{2} & -\phi & \frac{1}{\phi} \\
 \sqrt{2} & \frac{1}{\phi} & -\phi
\end{array}
\right)$ &${\tt diag}(0,1,4)$ \\
& &\\[-0.3cm]
\hline
& &\\[-0.3cm]
$3'$&
$-\frac{1}{\sqrt{5}}\left(
\begin{array}{ccc}
 1 & \sqrt{2} & \sqrt{2} \\
 \sqrt{2} & \frac{1}{\phi} & -\phi \\
 \sqrt{2} & -\phi & \frac{1}{\phi}
\end{array}
\right)$&${\tt diag}(0,2,3)$\\
& &\\[-0.3cm]
\hline
& &\\[-0.3cm]
$4$&$-\frac{1}{5}
\left(\begin{array}{cccc}
 -\sqrt{5} & (\phi-3) & (\phi+2) & -\sqrt{5} \\
 (\phi-3) & \sqrt{5} & \sqrt{5} & (\phi+2) \\
 (\phi+2) & \sqrt{5} & \sqrt{5} & (\phi-3) \\
 -\sqrt{5} & (\phi+2) & (\phi-3) & -\sqrt{5}
\end{array}
\right)$&${\tt diag}(1,2,3,4)$\\
& &\\
\hline
& &\\[-0.3cm]
$5$&$\frac{1}{5}\left(
\begin{array}{ccccc}
 -1 & \sqrt{6} & -\sqrt{6}  & -\sqrt{6}  & -\sqrt{6}  \\
  \sqrt{6}  & 2-\phi & 2\phi  & 2(1-\phi) & -1-\phi   \\
 -\sqrt{6}  & 2\phi  & 1+\phi & 2-\phi  & 2(-1+\phi) \\
 -\sqrt{6}  & 2(1-\phi)  & 2-\phi & 1+\phi & -2\phi  \\
 -\sqrt{6}  & -1-\phi   & -2(1-\phi)  & -2\phi & 2-\phi
\end{array}
\right)
$&${\tt diag}(0,1,2,3,4)$\\
& &\\
\hline
\end{tabular}
\caption{$S$ and $T$ generators of $A_5$ in the basis where $T$ is diagonal.}
\end{center}
\label{ca5}
\end{table}
\vskip 0.5cm
Notice that the matrices $S$ and $T$ of the previous section coincide with those of the $A_5$ generators in the representation 3.  
We have derived the Clebsh-Gordan coefficients entering the decomposition of the representation products. They are given in Appendix A.
\section{A model with $A_5$ family symmetry}
In this section we define our model. We focus on the lepton sector and, to facilitate the task
related to the vacuum alignment, we consider a supersymmetric model in the limit of exact supersymmetry (SUSY). SUSY breaking effects
do not affect lepton masses and mixing angles. Among the fields in the lepton sector we include three gauge singlets $\nu^c_i$ and the neutrino masses
will be dominated by the contribution of a type I see-saw mechanism \cite{SeeSaw}. A version of the model without see-saw, where the neutrino masses
are described by effective higher-dimensional operators, is equally possible. It would lead to the same predictions for the lepton
mixing angles. 

To start with we assign both the SU(2) lepton doublets $l$ and the right-handed neutrinos $\nu^c$ 
to the representation 3 of $A_5$. We take the SU(2) singlets $e^c$, $\mu^c$ and $\tau^c$ as invariant $A_5$ singlets. 
Higgs doublets $H_{u,d}$ are also singlets of $A_5$. 
In the neutrino sector we can write a renormalizable Yukawa coupling
of the type $(\nu^c l) H_u$, the notation (...) standing for the combination of the fields in parenthesis giving an $A_5$ singlet.
The product $\nu^c \nu^c$ is symmetric and contains a singlet and a pentaplet of $A_5$ and, to discuss the
most general case, we introduce two flavon chiral multiplets $\xi$ and $\varphi_S$ transforming as 1 and 5 of $A_5$, respectively.
They are completely neutral under the gauge interactions. In the charged lepton sector renormalizable Yukawa interactions
are not allowed and we need additional flavons transforming as 3 under $A_5$. To solve the vacuum alignment problem a minimum
of three triplets is needed, since trilinear interaction terms depending on less than three triplets vanish by the $A_5$ symmetry.
We include three triplets $\varphi$, $\varphi'$ and $\varphi''$, neutral under the gauge interactions. An additional flavon $\xi'$, singlet of $A_5$, is also
introduced to implement the desired vacuum alignment. 
To avoid couplings of the flavon multiplets to the wrong sector we also need to enlarge the flavour symmetry. 
This is done by considering the group $G_f=A_5\times Z_5 \times Z_3$.
In Table 3 we collect the chiral supermultiplets and their transformation properties under $G_f$.
Notice that, at variance with other constructions based on flavour symmetries, we do not introduce the so-called driving fields.
\vskip 0.5cm
\begin{table}[h!]
\begin{center}
\begin{tabular}{|c|c|c|c|c|c||c||c|c|c|c|c|c|}
\hline
&$e^c$&$\mu^c$&$\tau^c$&$l$&$\nu^c$&$H_{u,d}$&$\varphi$&$\varphi'$&$\varphi''$&$\varphi_S$&$\xi$&$\xi'$\\
\hline
${\bf A_5}$&$1$&$1$&$1$&$3$&$3$&$1$&$3$&$3$&$3$&$5$&$1$&$1$\\
\hline
\hline
$Z_5$&$0$&$4$&$1$&$0$&$0$&$0$&$0$&$4$&$1$&$0$&$0$&$0$\\
\hline
$Z_3$&$1$&$1$&$1$&$2$&$1$&$0$&$0$&$0$&$0$&$1$&$1$&$2$\\
\hline
\end{tabular}
\caption{Chiral multiplets and their transformation properties.}
\end{center}
\label{ca5}
\end{table}
The additional symmetry $Z_3$ is a discrete version of the total lepton number and is broken by the VEVs of the flavons of the neutrino sector,
$\varphi_S$, $\xi$ and $\xi'$. This symmetry prevents a direct mass term for $\nu^c$.
The presence of the new $Z_5$ factor forces each of the lepton multiplets $e^c$, $\mu^c$ and $\tau^c$ to
couple to only one of the triplets $\varphi$, $\varphi'$ and $\varphi''$, at the LO. The additional factors $Z_3$ and $Z_5$ play also an important role both in the construction of the flavon scalar potential 
and in the classification of NLO corrections. The superpotential for the lepton multiplets reads:
\vskip 0.5cm
\bea
w&=&y (\nu^c l) H_u+\dd\sqrt{\frac{3}{2}}y_1 \xi (\nu^c \nu^c)+y_5 (\varphi_S \nu^c \nu^c)\nn\\
&+&y_e e^c (\dd\frac{\varphi}{\Lambda} l) H_d+y_\mu \mu^c (\dd\frac{\varphi''}{\Lambda} l) H_d+y_\tau \tau^c (\dd\frac{\varphi'}{\Lambda} l) H_d+...
\label{w}
\eea
\vskip 0.5cm\noindent
where dots stand for higher order operators and $\Lambda$ denotes the cut-off scale. Notice that the LO Yukawa couplings
of the charged fermions are described by non-renormalizable operators. 
As we will see in section 5, where we will discuss the vacuum alignment, at the LO the flavons $\varphi_S$, $\xi$, $\varphi$, $\varphi'$ and $\varphi''$
acquire VEVs of the type:
\bea
\langle \varphi_S\rangle&=&(-\dd\sqrt{\frac{2}{3}}(p+q),-p,q,q,p)~\Lambda\nn\\
\langle \xi\rangle&=&s~\Lambda\nn\\
\langle\varphi \rangle&=&(u,0,0)~\Lambda\nn\\
\langle\varphi' \rangle&=&(0,u',0)~\Lambda\nn\\
\langle\varphi'' \rangle&=&(0,0,u'')~\Lambda~~~.
\label{vevs}
\eea
where $p,q,s,u,u',u''$ are dimensionless coefficients. Such a pattern completely specifies lepton masses and mixing angles, at the LO.
Plugging the VEVs of $\varphi_S$, $\xi$, $\varphi$, $\varphi'$ and $\varphi''$ into the superpotential $w$ and working out the $A_5$ invariant combinations,
with the help of the results of the previous section and those of the Appendix A, we can find the LO mass matrices $m_l$ and $m_\nu$.

In the charged lepton sector, after breaking of $A_5$, the relevant part of the superpotential becomes
\be
y_e u~ e^c l_e H_d+y_\mu u''~ \mu^c l_\mu H_d+y_\tau u'~ \tau^c l_\tau H_d~~~.
\ee
There is no contribution to the lepton mixing from this sector and charged lepton masses are 
\be
m_e=y_e u v_d~~~,~~~~~~~m_\mu=y_\mu u'' v_d~~~,~~~~~~~m_\tau=y_\tau u' v_d~~~,
\ee
$v_d$ being the VEV of the neutral component of $H_d$.
We might be surprised by the fact that $m_l$ is diagonal, since only the VEV of $\varphi$ leaves the $Z_5^{T}$ subgroup generated by $T$ invariant,
while the VEVs of $\varphi'$ and $\varphi''$ break $Z_5^{T}$. We can understand this result by recalling that the flavour symmetry $G_f$ contains
a factor $Z_5$, distinct from $Z_5^{T}$. The VEVs of $\varphi$, $\varphi'$ and $\varphi''$ break $A_5\times Z_5$ down to the
diagonal subgroup $Z_5^{D}$ contained in the product $Z_5^{T}\times Z_5$. It is this residual group that guarantees a diagonal $m_l$ in our construction.

Similarly, in the neutrino sector we read from eq. (\ref{w}) the mass matrices $M$ for the right-handed neutrinos and $m_D$, the Dirac one:
\be
M=\sqrt{6}\left(
\begin{array}{ccc}
y_1 s+\dd\frac{2}{3}y_5(p+q)&
y_5\dd\frac{p}{\sqrt{2}}&y_5\dd\frac{p}{\sqrt{2}}\\
y_5\dd\frac{p}{\sqrt{2}}&y_5 q&y_1 s-\dd\frac{1}{3}y_5(p+q)\\
y_5\dd\frac{p}{\sqrt{2}}&y_1 s-\dd\frac{1}{3}y_5(p+q)&y_5 q
\end{array}
\right)\Lambda~~~.
\ee
\be
m_D=y 
\left(
\begin{array}{ccc}
1&0&0\\
0&0&1\\
0&1&0
\end{array}
\right)
v_u~~~,
\ee
where $v_u$ is the VEV of the Higgs doublet $H_u$. Since $M$ is $\mu-\tau$ symmetric and $m_D$ is proportional to the matrix $U$ in eq. (\ref{sandu}), 
from the see-saw formula we have
\be
m_\nu=m_D^TM^{-1} m_D= y^2 v_u^2 M^{-1}
\ee
We notice that $M$ has precisely the structure given in eqs. (\ref{gl1}) and (\ref{solarGR}) and therefore both $M$ and its inverse are diagonalized by the 
mixing matrix in eq. (\ref{2.1}) with $\tan\theta_{12}=1/\phi$:
\be
U_{GR}^T~m_{\nu}~U_{GR}={\tt diag}(m_1,m_2,m_3)~~~.
\label{numass2}
\ee
Therefore $U_{GR}$ represents the contribution to the lepton mixing coming from the neutrino sector,
as desired. This result crucially depends on the VEV of the flavon pentaplet $\varphi_S$, which will be derived from the minimization of the
scalar potential in section 5. We observe that such a VEV is left invariant by the action of the generator $S$,
as can be immediately checked by multiplying the 5$\times$5 matrix $S$ of Table 2 and the vector $(-\sqrt{2/3}(p+q),-p,q,q,p)$. This is the reason why the residual symmetry
of the neutrino sector contains the parity subgroup generated by $S$. The presence of the $\mu-\tau$ symmetry is more subtle. Indeed the 
generator $S$ of the 5 representation has three eigenvalues equal to one and the corresponding eigenvector can be parametrized as
\be
(-\sqrt{2/3}(p+q+r\phi),-p+r,q+2r\phi,q,p+r)~~~.
\ee
This is the most general VEV of $\varphi_S$ that leaves $S$ unbroken. It is easy to construct the corresponding neutrino mass matrix $m_\nu$
and check that in the general case, with $r\ne 0$, $m_\nu$ is not $\mu-\tau$ symmetric. In our model it is the minimization of the scalar potential 
that selects the vacuum with $r=0$, thus enforcing the $\mu-\tau$ symmetry. We will demonstrate this result in section 5.

The charged fermion masses depend on three sets of independent parameters, which do not display a manifest relative hierarchy.
It is easy to induce the correct hierarchy by assigning Froggatt-Nielsen U(1)$_F$ charges $2q$ and $q$ to $e^c$, and $\mu^c$, respectively \cite{FN}.
The spontaneous breaking of such U(1)$_F$ by the VEV of a scalar fields carrying a negative units of $F$ explains why
$y_e<<y_\mu<<y_\tau$.
\begin{figure}[h!]
\begin{center}
\includegraphics[width=10cm,height=8cm,angle=0]{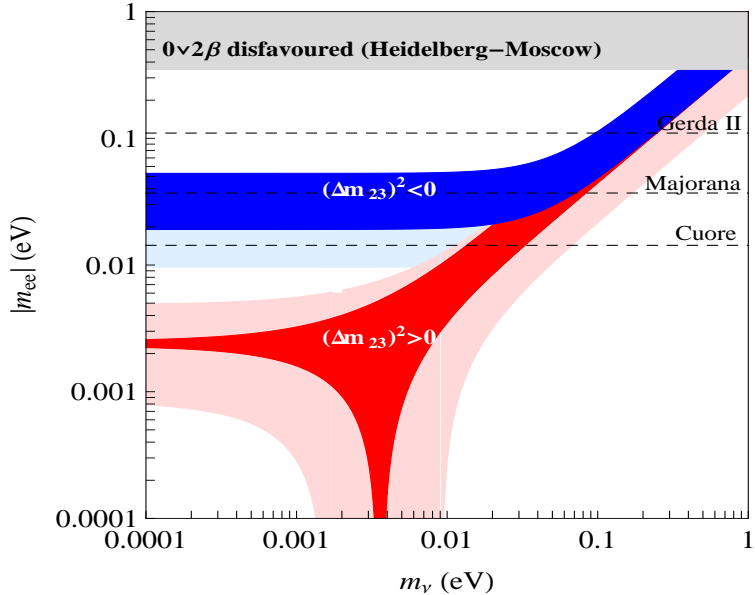}
\caption[]{Predictions for $|m_{ee}|$ versus the lightest neutrino mass, for normal (red) and inverted (blue) mass ordering.
In lighter regions the mass parameters are completely free. In darker regions they are correlated as prescribed by the LO approximation of 
our model.}

\label{fig:mee}
\end{center}
\end{figure}
In the LO approximation the spectrum of the light neutrinos is
\beq
m_1=\frac{1}{A+B+C}, \quad m_2=\frac{1}{A+B-C},\quad m_3=\frac{1}{-A+2 B}.
\eeq
where A, B and C are complex parameters defined as 
\beq
A= 6 y_1 q\frac{\Lambda}{\sqrt{6}y^2 v_u^2 }, \quad B = y_5(p+4q)\frac{\Lambda}{\sqrt{6} y^2 v_u^2} , \quad C= 3\sqrt{5} y_5 p \frac{\Lambda }{\sqrt{6} y^2 v_u^2}.
\eeq
There are no special relations between the three complex parameters and thus we have no prediction on the neutrino spectrum, that can have both normal and inverted ordering.
A moderate tuning among the parameters is needed in order to reproduce the 
ratio between solar and atmospheric squared mass differences.
The mass combination entering neutrinoless double-beta decay, $m_{ee}$, is given by
$y^2 v_u^2 (M^{-1})_{11}$ and depends on the same parameters $A$, $B$ and $C$.
By expressing the absolute values of $A$, $B$ and $C$ in terms of
$\Delta m^2_{sol}$, $\Delta m^2_{atm}$, the smallest neutrino mass and the phases of $A$, $B$ and $C$,
we can derive a range for $|m_{ee}|$ as a function of the smallest neutrino mass by varying
the available phases. We plot the result in figure 1, where the lighter region refer to the unconstrained case \cite{Strumia:review},
and the darker one corresponds to the present model, in the LO approximation. 

The dominant contribution from local effective operators to the light neutrino masses is $(l l h_u h_u \xi')$. This operator is 
suppressed compared to the see-saw contribution, since the former is of order  $VEV/\Lambda^2$, while the latter is of order $1/VEV$. Moreover,
it is easy to see that the new effective term preserves the $\mu-\tau$ symmetry and the Golden Ratio prediction. 

\section{Vacuum Alignment}
The results of the previous section crucially depend on the assumed set of VEVs, eq. (\ref{vevs}), and the purpose of this section
is to show that they derive from the minimization of the scalar potential of the theory, without ad-hoc tuning of the parameters involved.
The transformation properties of the flavon fields allow to write the following contribution to the superpotential
\bea
w_f&=&M_0 \xi \xi' +g_1 \xi (\varphi_S^2)+ g_2 (\varphi_S^3)_1+g_3 (\varphi_S^3)_2+\dd\frac{g_4}{3} \xi^3+\dd\frac{g_5}{3} \xi'^3\nn\\
&+&M_1 (\varphi^2)+M_2 (\varphi'\varphi'')+g(\varphi\varphi'\varphi'')+...
\eea
where dots stand for higher dimensional terms, which will contribute at the NLO. There are two independent cubic invariants
that can be built from a pentaplet and they are denoted by the suffices 1 and 2 in $w_f$.
There are no driving fields in our construction and the minima are derived by analyzing the
F-terms of the flavons themselves.
At the LO there is no mixing between $\xi$, $\xi'$, $\varphi_S$,
that control the neutrino mass terms and $\varphi$, $\varphi'$ and $\varphi''$, that give rise to the charged lepton Yukawas.
We can separately discuss the two sets of minima. We start from the neutrino sector. The condition 
\be
\dd\frac{\partial w_f}{\partial \xi'}=M_0 \xi+ g_5 \xi'^2=0
\ee
is solved by
\be
\xi=-\dd\frac{g_5}{M_0}\xi'^2~~~.
\label{sxi}
\ee
Another set of conditions is given by
\be
\dd\frac{\partial w_f}{\partial \varphi_{Si}}=0~~~.
\label{penta}
\ee
To solve these equations it is convenient to move to the so-called Cummins-Patera basis \cite{Cummins:1988,Ramond:2008} for the generators $S$ and $T$.
In this basis the generator $S$ for the five-dimensional representation is diagonal, $S_{CP}={\tt diag}(+1,-1,-1,+1,+1)$. The explicit form of $T$ for the 5 representation in the Cummins-Patera basis as well as the unitary matrix
relating the two basis is given in Appendix B. We denote the components of $\varphi_S$ in the Cummins-Patera basis by
\be
\varphi_S=(X_1,X_2,X_3,Z,\bar{Z})~~~,
\ee
where $X_i$ $(i=1,2,3)$, $Z$ and $\bar{Z}$ should be seen as independent complex quantities.
The terms of the superpotential $w_f$ that depend on $\varphi_S$ are explicitly given by:
\bea
w_f&=&g_1 \xi (X_1^2+X_2^2+X_3^2+2 Z\bar{Z})\nn\\
&+&g_2 \left(Z^3-\bar{Z}^3-3(X_1^2+\omega^2 X_2^2+\omega X_3^2) Z+3 (X_1^2+\omega X_2^2+\omega^2 X_3^2)\bar{Z}\right)\nn\\
&+&g_3\left(Z^3+\bar{Z}^3+(X_1^2+\omega^2 X_2^2+\omega X_3^2) Z+ (X_1^2+\omega X_2^2+\omega^2 X_3^2)\bar{Z}\right.\nn\\
&-&\left.4 X_1X_2X_3\right)+...
\eea
where 
\be
\omega=e^{\dd \frac{2 \pi i}{3}}~~~.
\ee
We have explicitly solved the equations (\ref{penta}) in this basis. We found no non-trivial solutions invariant under $T$ and
seven independent solutions invariant under $S$. They can be grouped in two pairs and a triplet.
Each of these sets is closed under the action of the generator $T$. One of the triplet of solutions is given by:
\be
X_1=X_2=X_3=0~~~,
\label{solx}
\ee
\bea
Z&=&-\dd\frac{2g_1 g_5}{3 M_0}\frac{1}{(g_2-g_3)^{1/3}(g_2+g_3)^{2/3}}\xi'^2~~~,\nn\\
{\bar Z}&=&+\dd\frac{2g_1 g_5}{3 M_0}\frac{1}{(g_2-g_3)^{2/3}(g_2+g_3)^{1/3}}\xi'^2~~~,
\label{mins}
\eea
where we have also made use of eq. (\ref{sxi}). The other two solutions belonging to the triplet
are obtained by multiplying $Z$ by $\omega(\omega^2)$ and $\bar{Z}$ by $\omega^2(\omega)$.
The condition $X_2=X_3=0$ correspond to the invariance under $S$, whereas $X_1=0$ is an additional specific feature of this set of solutions.

In each of these minima we have the basis independent result:
\be
(\varphi_S^2)=-\dd\frac{8 g_1^2 g_5^2}{9M_0^2}\frac{1}{(g_2-g_3)(g_2+g_3)}\xi'^4~~~.
\label{resind}
\ee
We have a finite multiplicity of minima and we chose the minimum in eqs. (\ref{solx},\ref{mins}). 
Before coming back to our basis, we analyze the equation
\be
\dd\frac{\partial w_f}{\partial \xi}=M_0 \xi'+ g_1 (\varphi_S^2)+g_4 \xi^2=0~~~,
\ee
which, by making use of (\ref{sxi}) and (\ref{resind}), becomes
\be
\dd\frac{\xi'}{M_0}- \left[\frac{8 g_1^3 g_5^2}{9(g_2-g_3)(g_2+g_3)} -g_4 g_5^2\right] \frac{\xi'^4}{M_0^4}=0~~~.
\ee
This equation has non-vanishing solutions for $\xi'$, which make non-trivial the solutions (\ref{sxi}) and (\ref{solx}, \ref{mins}). With the help of the unitary transformation relating the Cummins-Patera basis to ours we find that in our basis the minimum (\ref{solx}, \ref{mins}) translates into
\be
\langle \varphi_S\rangle=(-\dd\sqrt{\frac{2}{3}}(p+q),-p,q,q,p)~\Lambda
\label{alignment_pentaplet}
\ee
with
\bea
p&=&\frac{1}{2\sqrt{2}\Lambda}\left[\omega\left(\sqrt{\dd\frac{3}{5}}-i\right)Z-\left(\sqrt{\dd\frac{3}{5}}+i\right)\bar{Z}\right]\nn\\
q&=&\frac{1}{2\sqrt{2}\Lambda}\left[\omega\left(\sqrt{\dd\frac{3}{5}}+i\right)Z-\left(\sqrt{\dd\frac{3}{5}}-i\right)\bar{Z}\right]~~~,
\eea
with $Z$ and $\bar{Z}$ given by eq. (\ref{mins}). We have recovered the pattern displayed in 
section 4. Notice that this result does not depend on the specific value of the $Z$ and $\bar{Z}$
components, but rather on the conditions $X_i=0$. In particular, the $\mu-\tau$ parity symmetry
is related to the vanishing of the $X_1$ component. 

Moving to the flavons $\varphi$, $\varphi'$ and $\varphi''$, the relevant part of the superpotential is given by
\bea
w_f&=&M_1 (\varphi^2)+M_2 (\varphi'\varphi'')+g(\varphi\varphi'\varphi'')+...\nn\\
&=&M_1(\varphi_1^2+2 \varphi_2 \varphi_3)+M_2(\varphi'_1\varphi''_1+\varphi'_2\varphi''_3+\varphi'_3\varphi''_2)\nn\\
&+&g(\varphi_1\varphi'_2\varphi''_3+\varphi_2\varphi'_3\varphi''_1+\varphi_3\varphi'_1\varphi''_2-\varphi_1\varphi'_3\varphi''_2-\varphi_2\varphi'_1\varphi''_3-\varphi_3\varphi'_2\varphi''_1)+...
\label{LO}
\eea
The minima in the $\varphi$, $\varphi'$ and $\varphi''$ can be found by solving the system of equations:
\be
\dd\frac{\partial w_f}{\partial\varphi_i}=0~~~,~~~~~~~\dd\frac{\partial w_f}{\partial\varphi'_i}=0~~~,~~~~~~~\dd\frac{\partial w_f}{\partial\varphi''_i}=0~~~,
\ee
which, using a vectorial notation, can be written as
\be
2 M_1~ \hat{\varphi}+g~ \varphi'\times\varphi''=0~~~,~~~~
2 M_2~ \hat{\varphi}''-g~ \varphi\times\varphi''=0~~~,~~~~
2 M_2~ \hat{\varphi}'+g~ \varphi\times\varphi'=0~~~,
\label{system}
\ee
where $\times$ denotes the external product and, for any vector $v=(v_1,v_2,v_3)$, we set $\hat{v}=(v_1,v_3,v_2)$. 
To solve this system it is useful to recognize that the LO part of the superpotential that depends only on the fields $\varphi$, $\varphi'$ and $\varphi''$ is invariant  
under the linear trasformation
\be
\varphi\to \Omega  R~\Omega^{-1} \varphi~~~,~~~~~~~\varphi'\to y~\Omega R~\Omega^{-1} \varphi'~~~,~~~~~~~\varphi''\to \frac{1}{y}~\Omega R~\Omega^{-1}\varphi''~~~,
\label{rotate}
\ee
 where $y$ is a complex dimensionless parameter,
 \be
 \Omega=\left(
 \begin{array}{ccc}
 1&0&0\\
 & &\\[-1.0cm]\cr
 0&\dd\frac{1}{\sqrt{2}}&\dd\frac{-i}{\sqrt{2}}\\
 & &\\[-1.0cm]\cr
 0&\dd\frac{1}{\sqrt{2}}&\dd\frac{+i}{\sqrt{2}}
 \end{array}
 \right)
 \label{omega}
 \ee
 \vskip 0.3cm\noindent
and $R$ is a general complex orthogonal matrix, $R^T R=1$, depending on three complex parameters. As we shall see this invariance is accidental and is broken by the NLO contributions to the superpotential.
By exploiting such an invariance we can always reach the particular minimum with $\varphi_2=\varphi_3=0$. It is easy to see that one such solution is given by\footnote{There is also another solution where the entries of $\varphi'$ and $\varphi''$ are exchanged.}
\bea
\varphi_0&=&-\dd\frac{M_2}{g}\left(1,0,0\right)\nn\\
\varphi'_0&=&\dd\frac{\sqrt{2 M_1M_2}}{g} \left(0,1,0\right)\nn\\
\varphi''_0&=&\dd\frac{\sqrt{2M_1M_2}}{g }\left(0,0,1\right)~~~,
\label{solphip}
\eea
The general solution of the system ({\ref{system}) is given by
\be
\varphi= \Omega  R~\Omega^{-1} \varphi_0~~~,~~~~~~~\varphi'= y~\Omega R~\Omega^{-1} \varphi'_0~~~,~~~~~~~\varphi''=\frac{1}{y}~\Omega R~\Omega^{-1}\varphi'_0~~~,
\label{rotate0}
\ee
The degeneracies related to $R$ and $y$ are 
accidental. Indeed the transformations of eq. (\ref{rotate}) are not symmetries of our system, but rather accidental symmetries of the LO approximation.
It is easy to see that the symmetry related to the rotation $R$ is removed al the NLO, by including operators of dimension four in $w_f$. As we will discuss in the next section,
the inclusion of the most general set of operators of dimension four in the flavon fields, leads to the result $R=1$, thus justifying the choice of the previous section.
The symmetry under the rescaling $y$ is removed by adding invariant operators of dimension five.
\section{Higher-order corrections}

\subsection{Vacuum alignment}
The vacuum alignment discussed in the previous section is modified by
the contribution to the superpotential from higher dimensional operators.
If we denote by $VEV$ the typical vacuum expectation value of the flavon fields,
we expect corrections to the LO minima of order $VEV/\Lambda$.
These corrections can be kept small by asking $VEV/\Lambda\ll 1$.
Nevertheless they play an important role in removing some of the degeneracy
that affect the LO result. In the following discussion we include all NLO
operators, that is operators of dimension four depending on the flavon fields.
A complete set of invariants under the flavour group is given by
\be
\begin{array}{lcl}
Q_1=(\varphi \varphi)(\varphi \varphi)&&
Q_{11}=((\varphi \varphi)_5(\varphi' \varphi'')_5)\\
Q_2= (\varphi \varphi')(\varphi \varphi'')&&
Q_{12}=((\varphi' \varphi')_5(\varphi'' \varphi'')_5)\\
Q_3=(\varphi' \varphi'')(\varphi' \varphi'')&&
Q_{13}=(((\varphi_S\varphi_S)_{5_1}\varphi_S)_3\varphi)\\
Q_4=(\varphi' \varphi')(\varphi'' \varphi'')&&
Q_{14}=(((\varphi_S\varphi_S)_{5_2}\varphi_S)_3\varphi)\\
Q_5=(\varphi \varphi)(\varphi'\varphi'')&&
Q_{15}=(((\varphi_S\varphi_S)_4\varphi_S)_3\varphi)\\
Q_6=((\varphi \varphi')_3(\varphi \varphi'')_3)&&
Q_{16}=\xi'(\varphi_S(\varphi\varphi)_5)\\
Q_7=((\varphi' \varphi'')_3(\varphi' \varphi'')_3)&&
Q_{17}=\xi'(\varphi_S(\varphi'\varphi'')_5)\\
Q_8=((\varphi \varphi')_5(\varphi \varphi'')_5)&&
Q_{18}=\xi\xi'(\varphi\varphi)\\
Q_9=((\varphi' \varphi'')_5(\varphi' \varphi'')_5)&&
Q_{19}=\xi\xi'(\varphi'\varphi'')\\
Q_{10}=((\varphi \varphi)_5(\varphi \varphi)_5)&&
Q_{20}=\xi'^2(\varphi_S\varphi_S)~~~.
\end{array}
\label{NLO}
\ee
the NLO contribution to the flavon superpotential is
\be
\delta w_1+\delta w_2+\delta w_3
\ee
where
\be
\delta w_1=\sum_{i=1}^{12} x_i \dd\frac{Q_i}{\Lambda}~~~,~~~~~~~
\delta w_2=\sum_{i=13}^{19} x_i \dd\frac{Q_i}{\Lambda}~~~,~~~~~~~
\delta w_3= x_{20} \dd\frac{Q_{20}}{\Lambda}
\ee
It is useful to deal with the contribution $w_f+\delta w_1$ first. This includes all quartic operators that depend on the fields $\varphi$, $\varphi'$ and $\varphi''$ only. In this case 
the minima of $\varphi$, $\varphi'$ and $\varphi''$ can be analyzed in a analytic form. This part of the superpotential breaks the invariance of eq. (\ref{rotate}), but is still 
invariant  under the linear trasformation
\be
\varphi\to \Omega  R_{23}~\Omega^{-1} \varphi~~~,~~~~~~~\varphi'\to y~ \Omega R_{23}~\Omega^{-1} \varphi'~~~,~~~~~~~\varphi''\to\frac{1}{y}~\Omega R_{23}~\Omega^{-1}\varphi''~~~,
\label{rotate23}
\ee
with $\Omega$ given in eq. (\ref{omega}) and 
\be
R_{23}=\left(
\begin{array}{ccc}
1&0&0\\
0&\cos\alpha&\sin\alpha\\
0&-\sin\alpha&\cos\alpha
\end{array}
\right)
\ee
with $\alpha$ complex. We start by looking for a minimum for $\varphi$, $\varphi'$ and $\varphi''$ with the same orientation of the one in eq. (\ref{solphip})
\bea
\varphi_0&=&\left(u,0,0\right) \Lambda\nn\\
\varphi'_0&=&\left(0,u',0\right) \Lambda\nn\\
\varphi''_0&=&\left(0,0,u''\right)\Lambda ~~~,
\label{solphiNLO}
\eea
Along this direction the minimum conditions reduce to 
\bea
&&g u'u''+2 \frac{M_1}{\Lambda} u+4 (x_2+x_{11})u^3+ \dd\frac{1}{2}(4x_6-4x_7+3x_9+2x_{12}) u u' u''=0\nn\\
&&g u+\frac{M_2}{\Lambda}+\dd\frac{1}{4}(4x_6-4x_7+3x_9+2x_{12})u^2+\dd\frac{1}{2}(4x_4+4x_8+x_{10}+6x_{13})u'u''=0\nn~~~.
\eea
We find that the values of the $u$, $u'$ and $u''$ components are the ones given in eq. (\ref{solphip}) plus small perturbations of order $VEV/\Lambda$.
The solution (\ref{solphiNLO}) is not isolated. It is continuously connected to an infinite set of solutions given by
\be
\varphi= \Omega  R_{23}~\Omega^{-1} \varphi_0~~~,~~~~~~~\varphi'=y~ \Omega R_{23}~\Omega^{-1} \varphi'_0~~~,~~~~~~~\varphi''=\frac{1}{y}\Omega R_{23}~\Omega^{-1}\varphi'_0~~~.
\label{rotate0sol}
\ee
Thus the degeneracy present at the LO has been only partially removed by the NLO contribution $\delta w_1$. The remaining degeneracy is removed by the contribution $\delta w_2$.
We have analyzed the full NLO superpotential $w_f+\delta w_1+\delta w_2+\delta w_3$ by looking for numerical solution to the minimum equations. Looking for minima 
for the fields  $\varphi$, $\varphi'$ and $\varphi''$ we have frozen the value of $\varphi_S$ to its LO minimum, eq. (\ref{vevs}).
Under this condition it is easy to see that the operators $Q_{13}$, $Q_{14}$ and $Q_{15}$ vanish.
Moreover the operator $Q_{20}$ does not influence the minima of $\varphi$, $\varphi'$ and $\varphi''$ and the effect of the operators $Q_{18}$ and $Q_{19}$ can be absorbed in a redefinition 
of the parameters $M_1$ and $M_2$ of the LO superpotential. 
In our numerical simulation $g$, $x_{1-12}$ and $x_{16,17}$ are complex random numbers generated with a flat distribution in the square defined by the corners 
$[-(1+i)/\sqrt{2},(1+i)/\sqrt{2}]$. To get an expansion parameter $VEV/\Lambda$ of order 0.01, we have taken values of  $M_{1,2}/\Lambda$,  $\langle\varphi_S\rangle/\Lambda$ and $\xi'/\Lambda$ in the square defined by the corners $[-(1+i)/\sqrt{2},(1+i)/\sqrt{2}]\times 10^{-2}$. We have performed 50.000 independent minimizations of the scalar potential. We find that the mean values of the VEVs, normalized to one up to terms of order $(VEV/\Lambda)^2$, are  
\bea 
\langle\varphi\rangle&=&(100, 0.35+0.17 i, 0.35+0.17 i)\times 10^{-2}\nn\\
\langle\varphi'\rangle&=&(-0.51-0.03 i,100,-0.53-0.05i)\times 10^{-2}\nn\\
\langle\varphi''\rangle&=&(-0.51-0.03i,-0.53-0.05 i, 100)\times 10^{-2}~~~.
\label{triplet_mean}
\eea
Notice that the induced perturbations are not independent. We have $\langle\varphi_2\rangle=\langle\varphi_3\rangle$,
$\langle\varphi'_1\rangle=\langle\varphi''_1\rangle$ and $\langle\varphi'_3\rangle=\langle\varphi''_2\rangle$. This is true not only on average, but also
separately for each individual minimization.
The direction of the minima in flavour space is now completely determined and coincides,
up to corrections of relative order $VEV/\Lambda$ with the alignment (\ref{vevs}) needed to
enforce the desired mixing pattern. The only remaining flat direction is that related to the
overall scale of $\varphi'$ and $\varphi''$ (the parameter $y$ in eq. (\ref{rotate})), since also the NLO superpotential only depends on the combination $\varphi' \varphi''$. This last
flat direction is removed at NNLO order where terms depending separately on $\varphi'$ and $\varphi''$ first occur in the superpotential. 
Finally the contribution $\delta w_2$ also modifies the VEVs of $\xi$, $\xi'$ and $\varphi_S$ compared to their LO values. Also these corrections are of relative order $VEV/\Lambda$.

In summary the analysis of the scalar potential of the model in the SUSY limit shows that
the minima of the flavon fields are given by
\bea
\langle \varphi_S\rangle&=&(-\dd\sqrt{\frac{2}{3}}(p+q),-p,q,q,p)~\Lambda+O(\frac{VEV^2}{\Lambda})\nn\\
\langle \xi\rangle&=&s~\Lambda+O(\frac{VEV^2}{\Lambda})\nn\\
\langle\varphi \rangle&=&(u,0,0)~\Lambda+O(\frac{VEV^2}{\Lambda})\nn\\
\langle\varphi' \rangle&=&(0,u',0)~\Lambda+O(\frac{VEV^2}{\Lambda})\nn\\
\langle\varphi'' \rangle&=&(0,0,u'')~\Lambda+O(\frac{VEV^2}{\Lambda})~~~.
\label{vevs2}
\eea
This proves that the lepton mixing pattern originates from the dynamics of our model 
and not from an ad hoc choice of the underlying parameters.
  
\subsection{Other higher-order operators}
Beyond the operators (\ref{NLO}), that correct the lepton mass spectrum through the
flavon VEVs, there are other higher-dimensional operators contributing
directly to lepton masses. We only consider NLO contributions. 
At this order the charged lepton mass matrix $m_l$ is not affected. At LO $m_l$ is dominated
by operators of order $1/\Lambda$. At NLO we find the following
three invariant operators:
\bea
\frac{1}{\Lambda^2}e^c (\varphi' \varphi'' l)H_d, \nonumber\\
\frac{1}{\Lambda^2}\mu^c (\varphi'' \varphi l)H_d, \nonumber\\
\frac{1}{\Lambda^2}\tau^c(\varphi \varphi' l)H_d. 
\eea  
From the multiplication rules reported in the Appendix and the alignment shown in the previous sections it is easy to show that the NLO VEVs of the new triplets are 
\bea
(\varphi' \varphi'')_3 = (u'\; u'',0,0)\Lambda^2, \nonumber \\
(\varphi \varphi')_3 = (0,u\; u',0)\Lambda^2, \nonumber \\
(\varphi'' \varphi)_3 = (0,0,u'' \; u)\Lambda^2
\eea
and that they exactly align in the same direction as $\varphi$, $\varphi'$ and $\varphi''$, respectively. Thus these operators do not modify the LO structure of $m_l$. 
We conclude that, at NLO,  the charged lepton mass matrix is only modified by
the corrections to vacuum alignment analyzed in section 6.1.

The neutrino sector receives corrections from the operators:
\bea
&&(\nu^c l \varphi)H_u\nn\\
&&\xi'^2 (\nu^c \nu^c)\nonumber \\
&&(\varphi \varphi_S \nu^c \nu^c).
\label{honu}
\eea
The first one modifies non-trivially the Dirac neutrino mass matrix $m_D$. The second one, after the breaking of the flavour symmetry, can be absorbed by redefining the coupling constant $y_1$. 
The third one changes the Majorana mass matrix $M$ for the heavy neutrinos.
The neutrino mass matrix $m_\nu$ receives two type of corrections at the NLO.
One coming from the modified vacuum for the flavon $\varphi_S$ and another one
from the operators (\ref{honu}).  The corrections to the entries of $\varphi_S$ are unrelated to each other and consequently slightly modify the vacuum alignment
shown in eq. (\ref{alignment_pentaplet}). 
Neutrino masses and mixing angles are modified by terms of relative order $VEV/\Lambda$.
The size of this correction is constrained by the agreement between the predicted and observed
value of $\theta_{12}$. Not to spoil the successful prediction of $\theta_{12}$, the
ratio $VEV/\Lambda$ should not exceed a few percent.
In figure \ref{fig:3sigma} we show the relation between $\sin^2\theta_{12}$ and $\sin^2\theta_{13}$ at the NLO order as a result of a numerical 
simulation with random parameters. The simulation takes into account all the corrections coming from eqs. (\ref{vevs2}) and (\ref{honu}).

\begin{figure}[h!]
\begin{center}
\includegraphics[width=10cm,height=7cm,angle=0]{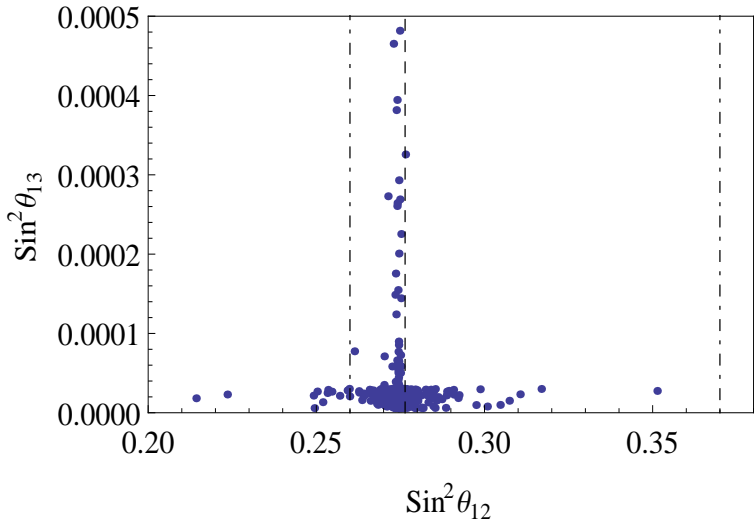}
\caption[]{Correlation between $\sin^2\theta_{12}$ and $\sin^2\theta_{13}$ at the NLO. For the triplets $\varphi$, $\varphi'$ and 
$\varphi''$ the mean values in (\ref{triplet_mean}) were chosen. $\langle\varphi_S\rangle/\Lambda$, $\langle\xi\rangle/\Lambda$ and $\langle\xi'\rangle/\Lambda$ have values 
in the square defined by the corners $[-(1+i)10^{-2}/\sqrt{2},(1+i)10^{-2}/\sqrt{2}]$. Similarly the other parameters are complex random 
numbers generated with a flat distribution in the square defined by the corners $[-(1+i)/\sqrt{2},(1+i)/\sqrt{2}]$. The LO
prediction for the solar angle is also shown (dashed line), along with the extremes of the $3\sigma$ experimental range (dot-dashed lines). }

\label{fig:3sigma}
\end{center}
\end{figure}
We recognize a possible correlation: in general either $\theta_{12}$ or $\theta_{13}$ can deviate significantly from the LO prediction, but
not both at the same time. When $VEV/\Lambda$ is of order 0.01, the maximal correction to the reactor angle remains far below the sensitivity of future experiments \cite{MezzettoSchwetz} and
the result from the global fit on neutrino oscillation given in \cite{bari}, where a value not too far from the current limit is reported.
On the other hand, $\theta_{12}$ could be shifted toward the experimental value shown in (\ref{exp_12}).

\section{Lepton mixing angles and Platonic Solids}
On a pure phenomenological basis there are attractive patterns that could provide a good LO approximation for the lepton mixing.
In particular the data are firmly indicating that the atmospheric mixing angle is close to maximal and that the reactor angle is the smallest one
so that, in a crude approximation, we can take $\sin^2 \theta_{23}=1/2$ and $\sin^2 \theta_{13}=0$.
In the same approximation, several choices have been suggested for the solar angle, such as $\sin^2 \theta_{12}=1/2$ (BM),
$\sin^2 \theta_{12}=1/3$ (TB) and $\tan \theta_{12}=1/\phi$ (GR), the choice of the present paper.
Clearly, depending on the pattern chosen as first approximation to the data, appropriate subleading corrections are needed, smaller for the TB and GR mixing patterns
and larger for the BM one.
In the basis where the charged leptons are diagonal the most general neutrino mass matrices corresponding to the considered LO approximations
are given by
\begin{equation}
m_\nu=\left(\matrix{
x&y&y\cr
y&z&w\cr
y&w&z}\right)~~~,
\label{gl}
\end{equation}
with coefficients $x$, $y$, $z$ and $w$ satisfying the following relations:
\bea
z+w&=&x~~~~~~~~~~~~~~{\tt BM}\cr
z+w&=&x+y~~~~~~~~~{\tt TB}\cr
z+w&=&x-\sqrt{2} y~~~~~{\tt GR}~~~.
\label{solar}
\eea
Thus we have three textures each depending on three independent complex parameters. 
Each of them can be completely characterized by a simple symmetry requirement, that of being invariant under two independent  commuting parity transformations, $U$ and $S$:
\be
S^2=U^2=1~~~,~~~~~~~~~~[S,U]=0~~~.
\ee
Indeed the mass matrix of eq. (\ref{gl}) is the most general one invariant under the so-called $\mu-\tau$ exchange symmetry, generated by:
\be
U=\left(
\begin{array}{ccc}
1&0&0\\
0&0&1\\
0&1&0
\end{array}
\right)~~~.
\label{mutau}
\ee
Moreover, depending on the particular chosen texture, we have another generator $S$, which can be found by using the constraints in eq. (\ref{solar}).
We list the solutions in Table 4.
The requirement of invariance under $U$ and $S$ completely determine the mass textures listed above. Namely, given a generic neutrino mass matrix $m_{\nu}$
the most general solutions to the equations:
\be
U^T~m_{\nu}~U=m_\nu~~~,~~~~~~~S^T~m_{\nu}~S=m_\nu~~~,
\ee
with $U$ and $S$ given in eq. (\ref{mutau}) and in Table 4, are the mass matrices defined by eqs. (\ref{gl}) and (\ref{solar}).

At the same time the requirement that the combination $m_l^\dagger m_l$ is diagonal can be fulfilled by asking that
the charged lepton sector is invariant under the action of a cyclic symmetry of $Z_n$ type. Calling $T$ the corresponding generator, 
with $T^n=1$, we can choose $T$ so that the solution of
\be
T^\dagger~(m^\dagger_l m_l)~T=(m^\dagger_l m_l)
\ee
is a diagonal matrix. In minimal constructions realizing BM, TB and GR mixings the generator $T$ can be chosen as in Table 4. With this choice $T$ has the
additional property:
\be
(ST)^3=1~~~.
\ee
\begin{table}[h!]
\begin{center}
\begin{tabular}{|c|c|c|c|}
\hline &\textbf{BM} & \textbf{TB} & \textbf{GR}\\
\hline 
& & &\\[-0.3cm]
$S$&$\dd\frac{1}{\sqrt{2}}\left(
    \begin{array}{ccc}
      0 & -1 & -1  \\
      -1 & 1/\sqrt{2} & -1/\sqrt{2} \\
      -1 & -1/\sqrt{2} & 1/\sqrt{2}\\
    \end{array}
  \right)$ & $\dd\frac{1}{3} \left(\matrix{
-1&2&2\cr
2&-1&2\cr
2&2&-1}\right)$ & 
$\dd\frac{1}{\sqrt{5}} \left(
\begin{array}{ccc}
1&\sqrt{2}&\sqrt{2}\\
\sqrt{2}&-\phi&1/\phi\\
\sqrt{2}&1/\phi&-\phi
\end{array}
\right)$
\\
& & &\\[-0.3cm]
\hline 
& & &\\[-0.3cm]
$T$&$\left(
    \begin{array}{ccc}
      -1 & 0 & 0  \\
      0 & -i & 0 \\
      0 & 0 & i\\
    \end{array}
  \right)$ & $ \left(\matrix{
1&0&0\cr
0&\omega&0\cr
0&0&\omega^2}\right)$ & 
$\left(
\begin{array}{ccc}
1&0&0\\
0&e^{\dd i\frac{2\pi}{5}}&0\\
0&0&e^{\dd i\frac{8\pi}{5}}
\end{array}
\right)$
\\
\hline
\end{tabular}
\caption{Generators S and  T for the different mixing patterns ($\phi=(1+\sqrt{5})/2$ and $\omega=-1/2+i \sqrt{3}/2$).}
\label{ts}
\end{center}
\end{table}

A model giving rise to the desired lepton mixing matrix $U_{PMNS}$ can be obtained starting from the family group $G_f$ generated by
$S$, $T$ and $U$. The family symmetry should be spontaneously broken by a set of scalar fields in such a way that, at the LO, the charged lepton
sector has a residual invariance under the group generated by $T$, whereas the neutrino sector has a residual invariance under
the group generated by $S$ and $U$. The desired lepton mixing $U_{PNMS}$ arises by construction, independently from the base choice.
In existing models the $\mu-\tau$ exchange symmetry generated by $U$ arises at the LO
as an accidental symmetry. In this case the family symmetry $G_f$ is generated by $S$ and $T$ satisfying:
\be
S^2=(ST)^3=1~~~~~~~~~~{\rm and}~~~~~~~~~~T^n=1~~~,
\label{present}
\ee
with $n=4$ for BM, $n=3$ for TB  and $n=5$ for GR. These are the defining relations of $S_4$ (BM) \cite{BM}, $A_4$ (TB) \cite{A4} and $A_5$ (GR), which are the proper symmetry group of the
cube/octahedron, tetrahedron and dodecahedron/icosahedron, respectively. 
We find rather intriguing that Platonic solids and their symmetries are in a natural
correspondence with the most popular lepton mixing patterns. Notice that these groups are all subgroups of the modular group, defined by
the presentation $S^2=(ST)^3=1$.

\section{Conclusion}
We think that the GR mixing pattern, where $\sin^2\theta_{23}=1/2$, $\theta_{13}=0$
and $\tan \theta_{12}=1/\phi$, should be considered on the same foot as other more popular schemes,
such as the TB and the BM ones, in our attempts to construct a model of  
lepton masses and mixing angles. Indeed the GR scheme is compatible with the experimental data. The largest deviation is for the solar angle, where the predicted value is about
two standard deviations below the present experimental central value. In this work we have built a supersymmetric model 
reproducing the GR pattern in the LO approximation. In the limit of exact GR mixing
we have identified two transformations $S$ and $T$, leaving invariant the neutrino mass matrix  and the charged lepton
mass matrix respectively, and generating the discrete group $A_5$. Following this hint, we have chosen as the family symmetry of our model
$A_5\times Z_5\times Z_3$ where the $Z_5\times Z_3$ factor forbids unwanted couplings between the flavon fields
and the matter fields. In the supersymmetric limit we have analyzed the most general scalar potential for the flavon fields
up to terms suppressed by one power of the cutoff $\Lambda$. In a finite portion of the parameter space, without any fine-tuning
of the parameters, we find an isolated minimum of the scalar potential where the flavon VEVs give rise to the GR
mixing pattern, up to terms of order $VEV/\Lambda$. Choosing $VEV/\Lambda$ of order few percent we can have 
an excellent agreement between theory and data for both the solar and the atmospheric mixing angles. The mixing angle
$\theta_{13}$ is expected to be of order few degrees. The neutrino masses depend on three complex parameters so that both types of ordering can be accommodated. Neutrino masses
and squared mass differences can be fitted but not predicted. At the LO we find restrictions on the allowed
value of $m_{ee}$, once the mass ordering and the smallest neutrino mass have been fixed.

To achieve the desired vacuum alignment there is no need of driving fields, a tool often used in this type of constructions. 
Neglecting matter multiplets, the energy density of the theory only depends on the flavon fields, that are self-aligned at the minimum.
To our best knowledge our model is the first example where the GR mixing pattern is derived from a full minimization
of the energy density.

We think that our model provides a valuable alternative to other existing proposals. There are not sufficient hints in the data to prefer
other mixing patterns, such as the TB one, to the GR one. Indeed in most of the existing models, including the present one, 
TB, BM, GR or other mixing patterns are only lowest order approximations, unavoidably corrected by powers of the 
symmetry breaking parameters. From this point of view, TB and GR mixing patterns can be both considered excellent first order approximations
to the existing data. It is remarkable that the TB, BM and GR mixing patterns can be obtained from minimal constructions
based on the symmetry groups $A_4$, $S_4$ and $A_5$, which are the proper symmetry groups of the Platonic solids.
To make a comparative experimental test of these constructions other observable quantities should be considered,
such as for instance the rates of lepton flavour violating processes, which, depending on the assumed supersymmetry breaking scale,
could be within the reach of the presently running or planned  experiments.
%
%
\section*{Acknowledgments}
We thank Claudia Hagedorn for useful comments and for pointing out several typos in the list of Clebsch-Gordan of the first version of our work. We recognize that this work has been partly supported by the European Commission under contract MRTN-CT-2006-035505 and by the European Programme "Unification in the LHC Era", contract PITN-GA-2009-237920 (UNILHC).
\section*{Appendix A. Kronecker products}
We report here the complete list of the Kronecker products for the group $A_5$. We assigne $a=(a_1,a_2,a_3)^T$ and $b=(b_1,b_2,b_3)^T$ 
to the {\bf 3} representation, while $a'=(a_1',a_2',a_3')^T$ and $b'=(b_1',b_2',b_3')^T$ belong to the ${\bf 3'}$ representation. 
$c=(c_1,c_2,c_3,c_4,c_5)^T$ and $d=(d_1,d_2,d_3,d_4,d_5)^T$ are pentaplets;  $f=(f_1,f_2,f_3,f_4)^T$ and $g=(g_1,g_2,g_3,g_4)^T$ 
are tetraplets.
\vspace{0.3cm}
\\
${\bf 3} \,\otimes\, {\bf 3}=~{\bf 3}_a~+~({\bf 1}~+~{\bf 5})_s\hfill$ 
\bea
{\bf 1}&=& a_1 b_1+a_2 b_3+a_3 b_2 \nonumber \\
{\bf 3}&=& (a_2 b_3 - a_3 b_2,~a_1 b_2 - a_2 b_1,~a_3 b_1-a_1 b_3)^T\nonumber \\
{\bf 5}&=& (a_1 b_1-\frac{a_2 b_3}{2}-\frac{a_3 b_2}{2},\frac{\sqrt{3}}{2}(a_1 b_2 + a_2 b_1), -\sqrt{\frac{3}{2}} a_2 b_2, -\sqrt{\frac{3}{2}} a_3 b_3,
	  -\frac{\sqrt{3}}{2}(a_1 b_3 + a_3 b_1))^T\nonumber
\eea 
\\
${\bf 3'}\,\otimes\, {\bf 3'}=~{\bf 3'}_a~+~({\bf 1}~+~{\bf 5})_s\hfill$ 
\bea
{\bf 1}&=& a'_1 b'_1+a'_2 b'_3+a'_3 b'_2 \nonumber \\
{\bf 3'}&=& (a'_2 b'_3 - a'_3 b'_2,~a'_1 b'_2 - a'_2 b'_1,~a'_3 b'_1-a'_1 b'_3)^T\nonumber \\
{\bf 5}&=& (a'_1 b'_1-\frac{a'_2 b'_3}{2}-\frac{a'_3 b'_2}{2},\sqrt{\frac{3}{2}} a'_3 b'_3,-\frac{\sqrt{3}}{2}(a'_1 b'_2 + a'_2 b'_1),-\frac{\sqrt{3}}{2}(a'_1 b'_3 + a'_3 b'_1),
	     -\sqrt{\frac{3}{2}} a'_2 b'_2)^T\nonumber
\eea
\\
$ {\bf 3}\,\otimes\, {\bf 3'} =~{\bf 4}~+~{\bf 5}\hfill$ 
\bea
{\bf 4}&=& (a_2 b'_1 - \frac{a_3 b'_2}{\sqrt{2}},-a_1 b'_2 + \frac{a_3 b'_3}{\sqrt{2}},a_1 b'_3 - \frac{a_2 b'_2}{\sqrt{2}},
	      -a_3 b'_1 + \frac{a_2 b'_3}{\sqrt{2}})^T\nonumber \\
{\bf 5}&=& (a_1 b'_1,-\frac{a_2 b'_1 + \sqrt{2}a_3 b'_2}{\sqrt{3}},\frac{a_1 b'_2 + \sqrt{2}a_3 b'_3}{\sqrt{3}},
	    \frac{a_1 b'_3 + \sqrt{2}a_2 b'_2}{\sqrt{3}},\frac{a_3 b'_1 + \sqrt{2}a_2 b'_3}{\sqrt{3}}))^T\nonumber
\eea
\\
${\bf 3}\,\otimes\, {\bf 4}\,~=~{\bf 3'}~+~{\bf 4}~+~{\bf 5}\hfill$ 
\bea
{\bf 3'}&=& (a_2 g_4 -a_3 g_1,\frac{1}{\sqrt{2}}(\sqrt{2} a_1 g_2 +a_2 g_1 +a_3 g_3),-\frac{1}{\sqrt{2}}(\sqrt{2} a_1 g_3 +a_2 g_2 +a_3 g_4))^T
	    \nonumber \\
{\bf 4}&=& (a_1 g_1 + \sqrt{2} a_3 g_2,-a_1 g_2 + \sqrt{2} a_2 g_1,a_1 g_3 - \sqrt{2} a_3 g_4,
	      -a_1 g_4 - \sqrt{2} a_2 g_3)^T\nonumber \\
{\bf 5}&=& (a_3 g_1 +a_2 g_4,\sqrt{\frac{2}{3}}(\sqrt{2} a_1 g_1-a_3 g_2),\frac{1}{\sqrt{6}}(\sqrt{2}a_1 g_2-3 a_3 g_3+a_2 g_1 ),\nonumber \\
	& &  \frac{1}{\sqrt{6}}(\sqrt{2}a_1 g_3-3 a_2 g_2+a_3 g_4 ),\sqrt{\frac{2}{3}}(-\sqrt{2} a_1 g_4+a_2 g_3))^T\nonumber
\eea
\\
${\bf 3'}\,\otimes\, {\bf 4}\,~=~{\bf 3}~+~{\bf 4}~+~{\bf 5}\hfill$ 
\bea
{\bf 3}&=& (a'_2 g_3 -a'_3 g_2,\frac{1}{\sqrt{2}}(\sqrt{2} a'_1 g_1 +a'_2 g_4-a'_3 g_3),\frac{1}{\sqrt{2}}(-\sqrt{2} a'_1 g_4 +a'_2 g_2 -a'_3 g_1))^T
	    \nonumber \\
{\bf 4}&=& (a'_1 g_1 + \sqrt{2} a'_3 g_3,a'_1 g_2 - \sqrt{2} a'_3 g_4,-a'_1 g_3 + \sqrt{2} a'_2 g_1,
	      -a'_1 g_4 - \sqrt{2} a'_2 g_2)^T\nonumber \\
{\bf 5}&=& (a'_3 g_2 +a'_2 g_3,\frac{1}{\sqrt{6}}(\sqrt{2}a'_1 g_1 -3 a'_2 g_4-a'_3 g_3 ),-\sqrt{\frac{2}{3}}(\sqrt{2} a'_1 g_2+a'_3 g_4),\nonumber \\
	& & -\sqrt{\frac{2}{3}}(\sqrt{2} a'_1 g_3+a'_2 g_1),\frac{1}{\sqrt{6}}(-\sqrt{2}a'_1 g_4+3 a'_3 g_1+a'_2 g_2 ))^T\nonumber
\eea
\\
\\
${\bf 3 }\,\otimes\, {\bf 5}\,~=~{\bf 3}~+~{\bf 3'}~+~{\bf 4}~+~{\bf 5}\hfill$ 
\bea
{\bf 3}&=&(\frac{2 a_1 c_1}{\sqrt{3}}+ a_3 c_2 -a_2 c_5,-\frac{ a_2 c_1}{\sqrt{3}}+ a_1 c_2 -\sqrt{2}a_3 c_3,
	  -\frac{ a_3 c_1}{\sqrt{3}}- a_1 c_5 -\sqrt{2}a_2 c_4)^T\nonumber \\
{\bf 3'}&=&( a_1 c_1+\frac{a_2 c_5-a_3 c_2}{\sqrt{3}},\frac{ a_1 c_3 +\sqrt{2}(a_3 c_4-a_2 c_2)}{\sqrt{3}},
	  \frac{ a_1 c_4 +\sqrt{2}(a_2 c_3 + a_3 c_5)}{\sqrt{3}})^T\nonumber\\
{\bf 4}&=&(4 a_1 c_2 +2\sqrt{3}a_2 c_1 +\sqrt{2}a_3 c_3,2 a_1 c_3 -2\sqrt{2}a_2 c_2 -3\sqrt{2}a_3 c_4,\nonumber \\
	& &    2 a_1 c_4 -3\sqrt{2}a_2 c_3 +2 \sqrt{2}a_3 c_5,-4 a_1 c_5 +\sqrt{2}a_2 c_4 +2\sqrt{3}a_3 c_1)^T \nonumber \\
{\bf 5}&=&(a_2 c_5 +a_3 c_2,a_2 c_1 -\frac{a_1 c_2 +\sqrt{2}a_3 c_3}{\sqrt{3}},-\frac{2 a_1 c_3 +\sqrt{2}a_2 c_2}{\sqrt{3}},\nonumber \\
	& &    \frac{2 a_1 c_4 -\sqrt{2}a_3 c_5}{\sqrt{3}},a_3 c_1 +\frac{a_1 c_5 -\sqrt{2}a_2 c_4}{\sqrt{3}})^T\nonumber
\eea
\\
${\bf 3'}\,\otimes\, {\bf 5}\,~=~{\bf 3}~+~{\bf 3'}~+~{\bf 4}~+~{\bf 5}\hfill$ 
\bea
{\bf 3}&=&( a'_1 c_1+\frac{a'_3 c_3+a'_2 c_4}{\sqrt{3}},\frac{- a'_1 c_2 +\sqrt{2}(a'_3 c_4+a'_2 c_5)}{\sqrt{3}},
	  \frac{ a'_1 c_5 +\sqrt{2}(a'_2 c_3 - a'_3 c_2)}{\sqrt{3}})^T\nonumber\\
{\bf 3'}&=&(\frac{2 a'_1 c_1}{\sqrt{3}}- a'_3 c_3 -a'_2 c_4,-\frac{ a'_2 c_1}{\sqrt{3}}- a'_1 c_3 -\sqrt{2}a'_3 c_5,
	  -\frac{ a'_3 c_1}{\sqrt{3}}- a'_1 c_4 +\sqrt{2}a'_2 c_2)^T\nonumber \\
{\bf 4}&=&(2 a'_1 c_2 + 3\sqrt{2}a'_2 c_5 -2\sqrt{2}a'_3 c_4,-4 a'_1 c_3  +2\sqrt{3}a'_2 c_1 +\sqrt{2}a'_3 c_5,\nonumber \\
	& &    -4 a'_1 c_4 -\sqrt{2}a'_2 c_2 +2\sqrt{3}a'_3 c_1 ,-2 a'_1 c_5 -2\sqrt{2}a'_2 c_3 - 3\sqrt{2}a'_3 c_2)^T \nonumber \\
{\bf 5}&=&(a'_2 c_4 -a'_3 c_3,\frac{2 a'_1 c_2 +\sqrt{2}a'_3 c_4}{\sqrt{3}},-a'_2 c_1 -\frac{a'_1 c_3 -\sqrt{2}a'_3 c_5}{\sqrt{3}},\nonumber \\
	& &  a'_3 c_1 +\frac{a'_1 c_4 +\sqrt{2}a'_2 c_2}{\sqrt{3}},\frac{-2 a'_1 c_5 +\sqrt{2}a'_2 c_3}{\sqrt{3}})^T\nonumber
\eea
\\
$~\,{\bf 4}\,\otimes\, {\bf 4}\,~=~({\bf 3}~ +~{\bf 3'})_a~+~({\bf 1}~+~{\bf 4}~ +~{\bf 5})_s\hfill$ 
\bea
{\bf 1}&=& f_1 g_4 +f_2 g_3 +f_3 g_2 +f_4 g_1 \nonumber \\
{\bf 3}&=&(f_1 g_4 -f_4 g_1 +f_3 g_2- f_2 g_3,\sqrt{2}(f_2 g_4 -f_4 g_2), \sqrt{2}(f_1 g_3 -f_3 g_1))^T\nonumber \\
{\bf 3'}&=&(f_1 g_4 -f_4 g_1 +f_2 g_3- f_3 g_2,\sqrt{2}(f_3 g_4 -f_4 g_3), \sqrt{2}(f_1 g_2 -f_2 g_1))^T\nonumber \\
{\bf 4}&=&(f_3 g_3 -f_4 g_2 -f_2 g_4,f_1 g_1 +f_3 g_4 +f_4 g_3, -f_4 g_4 -f_1 g_2 -f_2 g_1,-f_2 g_2 +f_1 g_3 +f_3 g_1)^T\nonumber \\
{\bf 5}&=&(f_1 g_4 +f_4 g_1 -f_3 g_2 -f_2 g_3,-\sqrt{\frac{2}{3}}(2 f_3 g_3 +f_2 g_4 +f_4 g_2), 
	    \sqrt{\frac{2}{3}}(-2 f_1 g_1 +f_3 g_4 +f_4 g_3),\nonumber \\
	& &    \sqrt{\frac{2}{3}}(-2 f_4 g_4 +f_2 g_1 +f_1 g_2),\sqrt{\frac{2}{3}}(2 f_2 g_2 +f_1 g_3 +f_3 g_1) )^T\nonumber 
\eea
\\
$~\,{\bf 4}\,\otimes\, {\bf 5}\,~=~{\bf 3}~ +~{\bf 3'}~+~{\bf 4}~+~{\bf 5}~ +~{\bf 5}\hfill$ 
\bea
{\bf 3}&=&(4 f_1 c_5 -4 f_4 c_2 -2 f_3 c_3-2 f_2 c_4,-2 \sqrt{3} f_1 c_1-\sqrt{2}(2 f_2 c_5 -3 f_3 c_4+f_4 c_3),\nonumber\\ 
	& &     \sqrt{2} (-f_1 c_4+ 3 f_2 c_3 +2 f_3 c_2)-2\sqrt{3} f_4 c_1)^T\nonumber \\
{\bf 3'}&=&(2 f_1 c_5 -2 f_4 c_2 +4 f_3 c_3+4 f_2 c_4,-2 \sqrt{3} f_2 c_1+\sqrt{2}(2 f_4 c_4 +3 f_1 c_2-f_3 c_5), \nonumber\\
	& &     \sqrt{2} (f_2 c_2- 3 f_4 c_5 +2 f_1 c_3)-2\sqrt{3} f_3 c_1)^T\nonumber \\
{\bf 4}&=&(3 f_1 c_1+\sqrt{6}(f_2 c_5 +f_3 c_4-2 f_4 c_3),-3 f_2 c_1+\sqrt{6}(f_4 c_4 -f_1 c_2 +2 f_3 c_5),\nonumber\\
	& &    -3 f_3 c_1+\sqrt{6}(f_1 c_3 +f_4 c_5-2 f_2 c_2),3 f_4 c_1+\sqrt{6}(f_2 c_3 -f_3 c_2-2 f_1 c_4))^T\nonumber \\
{\bf 5_1}&=&(f_1 c_5 +2 f_2 c_4 -2 f_3 c_3+f_4 c_2,-2 f_1 c_1+\sqrt{6} f_2 c_5 ,f_2 c_1+\sqrt{\frac{3}{2}}(-f_1 c_2 -f_3 c_5+2 f_4 c_4),\nonumber\\
	& &    -f_3 c_1-\sqrt{\frac{3}{2}}(f_2 c_2 +f_4 c_5+2 f_1 c_3),-2 f_4 c_1-\sqrt{6} f_3 c_2)^T\nonumber \\
{\bf 5_2}&=&(f_2 c_4 - f_3 c_3,-f_1 c_1+\frac{2 f_2 c_5-f_3 c_4 -f_4 c_3}{\sqrt{6}},-\sqrt{\frac{2}{3}}(f_1 c_2 +f_3 c_5- f_4 c_4),\nonumber\\
	& &    -\sqrt{\frac{2}{3}}(f_1 c_3 +f_2 c_2+ f_4 c_5),-f_4 c_1-\frac{2 f_3 c_2+f_1 c_4 +f_2 c_3}{\sqrt{6}})^T\nonumber 
\eea
\\
$~\,{\bf 5}\,\otimes\, {\bf 5}\,~=~({\bf 3}~ +~{\bf 3'}~ +~{\bf 4})_a~ +~({\bf 1}~+~{\bf 4}~ +~{\bf 5}~ +~{\bf 5})_s\hfill$ 
\bea
{\bf 3}&=&(2 ( c_4 d_3 -c_3 d_4)+c_2 d_5 -c_5 d_2,\sqrt{3}(c_2 d_1-c_1 d_2)+\sqrt{2}(c_3 d_5 -c_5 d_3),\nonumber \\
	& &    \sqrt{3}(c_5 d_1-c_1 d_5)+\sqrt{2}(c_4 d_2 -c_2 d_4))^T\nonumber \\
{\bf 3'}&=&(2 ( c_2 d_5 -c_5 d_2)+c_3 d_4 -c_4 d_3,\sqrt{3}(c_3 d_1-c_1 d_3)+\sqrt{2}(c_4 d_5 -c_5 d_4),\nonumber \\
	& &  \sqrt{3}(c_1 d_4-c_4 d_1)+\sqrt{2}(c_3 d_2 -c_2 d_3))^T\nonumber \\
{\bf 4_s}&=& ((c_1 d_2 +c_2 d_1)-\frac{(c_3 d_5 +c_5 d_3) -4 c_4 d_4}{\sqrt{6}},-(c_1 d_3 + c_3 d_1)-\frac{(c_4 d_5 +c_5 d_4) -4 c_2 d_2}{\sqrt{6}},\nonumber \\
         & & (c_1 d_4 +c_4 d_1)-\frac{(c_2 d_3 +c_3 d_2) +4 c_5 d_5}{\sqrt{6}},(c_1 d_5 + c_5 d_1)-\frac{(c_2 d_4 +c_4 d_2) +4 c_3 d_3}{\sqrt{6}})^T \nonumber \\
{\bf 4_a}&=& ((c_1 d_2 -c_2 d_1)+\sqrt{\frac{3}{2}}(c_3 d_5-c_5 d_3),(c_1 d_3 - c_3 d_1)+\sqrt{\frac{3}{2}}(c_4 d_5-c_5 d_4),\nonumber \\
         & & (c_4 d_1 -c_1 d_4)+\sqrt{\frac{3}{2}}(c_3 d_2-c_2 d_3),(c_1 d_5 - c_5 d_1)+\sqrt{\frac{3}{2}}(c_4 d_2-c_2 d_4))^T \nonumber \\
{\bf 5_1}&=& (c_1 d_1 +c_2 d_5 +c_5 d_2 +\frac{c_3 d_4 +c_4 d_3}{2}, -(c_1 d_2 +c_2 d_1)+ \sqrt{\frac{3}{2}} c_4 d_4,
             \frac{1}{2}(c_1 d_3 + c_3 d_1-\sqrt{6}(c_4 d_5 +c_5 d_4)),\nonumber \\
         & & \frac{1}{2}(c_1 d_4 + c_4 d_1+\sqrt{6}(c_2 d_3 +c_3 d_2)),-(c_1 d_5 +c_5 d_1) -\sqrt{\frac{3}{2}} c_3 d_3)^T \nonumber \\
{\bf 5_2}&=& (\frac{2 c_1 d_1 +c_2 d_5 +c_5 d_2}{2}, \frac{-3(c_1 d_2+c_2 d_1)+\sqrt{6}(2 c_4 d_4 +c_3 d_5+c_5 d_3)}{6},
             -\frac{2 c_4 d_5 +2 c_5 d_4+c_2 d_2}{\sqrt{6}},\nonumber \\
         & & \frac{2 c_2 d_3 +2 c_3 d_2-c_5 d_5}{\sqrt{6}},\frac{-3(c_1 d_5+c_5 d_1)+\sqrt{6}(-2 c_3 d_3 +c_2 d_4+c_4 d_2)}{6})^T \nonumber \\
{\bf 1}&=& c_1 d_1+c_3 d_4+c_4 d_3-c_2 d_5 -c_5 d_2 \nonumber 
\eea
\newpage
\section*{Appendix B. Other representations of $A_5$}
{\bf Shirai's basis.}
In this section we show how the basis chosen in this work is related to the representations given in previous articles on the group $A_5$. 
In \cite{everett:2009,Shirai} the Shirai base was used, in which the presentation of the group is given in terms of two matrices $S_{Sh}$ and $T_{Sh}$ satisfying 
the following algebra:
\beq
S^2_{Sh} =T_{Sh}^5=(T^2_{Sh} S_{Sh} T_{Sh}^3 S_{Sh} T_{Sh}^{-1} S_{Sh} T_{Sh} S_{Sh} T_{Sh}^{-1})^3= 1
\eeq
In order to connect them to the matrix $S$ and $T$ satisfying 
\beq
S^2= T^5 =(ST)^3=1
\eeq
with $T$ diagonal, we take an intermediate step and define first
\beq
S'=S_{Sh}
\eeq
and a matrix T' such that $(S'T')^3=1$
Then, we define  $T^2_{Sh} S_{Sh} T_{Sh}^3 S_{Sh} T_{Sh}^{-1} S_{Sh} T_{Sh} S_{Sh} T_{Sh}^{-1}= A_{Sh}$. Since $S'T'= A_{Sh}$ we obtain
\beq
T'=S'^{-1} A_{Sh}=S' A_{Sh} =S_{Sh} A_{Sh}
\eeq
Note that $T'$ is not diagonal. To make it diagonal, we let an unitary matrix U act on it, such that
\beq
S=U^\dagger S' U, \qquad T=U^\dagger T' U.
\eeq
and finally
\beq
S=U^\dagger S_{Sh} U, \qquad T=U^\dagger S_{Sh} A_{Sh} U.
\eeq
{\bf Cummins-Patera's basis.}
The Cummins-Patera's basis \cite{Cummins:1988,Ramond:2008} is generated by two elements with presentation 
\be
A^2=B^3=(AB)^5=1,
\ee
and the quintic representation are explicitly given by
\be
A\;= \pmatrix{
\hfill 1&\hfill 0&\hfill 0&\hfill 0&\hfill 0\cr 
\hfill 0&\hfill -1&\hfill 0&\hfill 0&\hfill 0\cr
\hfill 0&\hfill 0&\hfill -1&\hfill 0&\hfill 0\cr
\hfill 0&\hfill 0&\hfill 0&\hfill 1&\hfill 0\cr
\hfill 0&\hfill 0&\hfill 0&\hfill 0&\hfill 1} ;
\quad
B\;=\;-\frac{1}{2}\pmatrix{1&0&1&-\omega^2&-\omega\cr
0&1&1&-1&-1\cr
-1&-1&0&\omega&\omega^2\cr
\omega&1&\omega^2&0&-\omega\cr
\omega^2&1&\omega&-\omega^2&0}, \label{AandB}
\ee
where $\omega=e^{\frac{2i\pi}{3}}$ is the cubic root of unity, 
We can define a unitary transformation $U_{CP}$ that relates the elements of the Cummins-Patera basis to the one introduced in Section 3 as 
follows:
\be
S=U_{CP}^\dagger A U_{CP},\quad (ST)=U_{CP}^\dagger B U_{CP},\quad T=U_{CP}^\dagger (A B) U_{CP}.
\ee
The unitary matrix is 
\be
U_{CP}\;= \frac{1}{\sqrt{5}}\pmatrix{
\hfill 0&\hfill \omega^{-\frac{1}{2}} \sqrt{\frac{\sqrt{5}}{2 \phi }} &\hfill  \sqrt{\frac{\sqrt{5}\phi}{2  }}&\hfill -\omega^{-\frac{1}{2}} \sqrt{\frac{\sqrt{5}\phi}{2  }}&\hfill \omega^{-\frac{1}{2}} \sqrt{\frac{\sqrt{5}}{2 \phi }}\cr 
\hfill -\sqrt{3}\omega^{\frac{1}{4}} &\hfill \frac{\omega^{\frac{1}{4}}}{\sqrt{2}}&\hfill -\frac{\omega^{\frac{1}{4}}}{\sqrt{2}}&\hfill -\frac{\omega^{\frac{1}{4}}}{\sqrt{2}}&\hfill -\frac{\omega^{\frac{1}{4}}}{\sqrt{2}}\cr
\hfill 0 &\hfill \omega^{-\frac{1}{2}} \sqrt{\frac{\sqrt{5}\phi}{2  }}&\hfill \omega^{-\frac{1}{2}} \sqrt{\frac{\sqrt{5}}{2 \phi }}&\hfill \omega^{-\frac{1}{2}} \sqrt{\frac{\sqrt{5}}{2 \phi }}&\hfill -\omega^{-\frac{1}{2}} \sqrt{\frac{\sqrt{5}\phi}{2  }}\cr
\hfill -\omega^2&\hfill -e^{i \alpha} &\hfill -e^{i \beta}&\hfill -e^{i \alpha}&\hfill e^{i \alpha}\cr
\hfill 1&\hfill e^{-i\gamma}&\hfill -e^{i\gamma}&\hfill -e^{i\gamma}&\hfill -e^{-i\gamma}},
\ee
where $\alpha=\arctan{(-\frac{2+\sqrt{5}}{\sqrt{3}})}$, $\beta=\arctan{(\sqrt{3-\frac{4\sqrt{5}}{3}})}$ and $\gamma=\arctan{(\sqrt{\frac{5}{3}})}$.
It is straightforward to verify that acting with $U^\dagger_{CP}$ on the vector $(0,0,0,Z,\overline{Z})^T$ gives a new vector with the form given in eq. (\ref{alignment_pentaplet}).

\end{document}